\documentclass[useAMS,usenatbib]{mnras}
\usepackage{natbib,amsbsy}
\usepackage{amsmath,amssymb}
\usepackage{graphicx}
\usepackage{subfigure}
\usepackage{textcomp}
\usepackage[T1]{fontenc}
\usepackage{ae,aecompl}
 \usepackage{dblfloatfix}
 \usepackage[normalem]{ulem}

\title[]{The Properties of Bound and Unbound Molecular Cloud Populations Formed in Galactic Disc Simulations}
\author[R.~L. Ward et al.]{Rachel L. Ward\thanks{Email: rlward@mcmaster.ca}, Samantha M. Benincasa, James Wadsley, Alison Sills, 
\newauthor
and H.~M.~P. Couchman \\
Department of Physics and Astronomy, McMaster University, Hamilton, ON, L8S 4M1, Canada} 

\begin{document}
\label{firstpage}
\pagerange{\pageref{firstpage}--\pageref{lastpage}} \pubyear{2015}

\maketitle
\begin{abstract}
	We explore the effect of galactic environment on properties of molecular clouds.  Using clouds formed in a large-scale galactic disc simulation, we measure the observable properties from synthetic column density maps.  We confirm that a significant fraction of unbound clouds forms naturally in a galactic disc environment and that a mixed population of bound and unbound clouds can match observed scaling relations and distributions for extragalactic molecular clouds.  By dividing the clouds into inner and outer disc populations, we compare their distributions of properties and test whether there are statistically significant differences between them.  We find that clouds in the outer disc have lower masses, sizes, and velocity dispersions as compared to those in the inner disc for reasonable choices of the inner/outer boundary.  We attribute the differences to the strong impact of galactic shear on the disc stability at large galactocentric radii.  In particular, our Toomre analysis of the disc shows a narrowing envelope of unstable masses as a function of radius, resulting in the formation of smaller, lower mass fragments in the outer disc.  We also show that the star formation rate is affected by the environment of the parent cloud, and is particularly influenced by the underlying surface density profile of the gas throughout the disc.  Our work highlights the strengths of using galaxy-scale simulations to understand the formation and evolution of cloud properties -- and the star formation within them -- in the context of their environment.
\end{abstract}


\section{Introduction}\label{sec:Intro4}

Stars form principally, and possibly exclusively, in giant molecular clouds (GMCs) which are large conglomerations of gas and dust, predominantly composed of molecular hydrogen and concentrated in the arms of spiral galaxies.  With the advent of powerful new telescopes like ALMA with the ability to reach previously unprecedented resolutions, it is now possible to study local clouds in greater detail than ever before.  However, limited information can be obtained on the internal structure of extragalactic GMCs from observations of nearby galaxies as they typically have cloud-scale resolution.  Observers can partially overcome this limitation by combining data from various surveys to probe the multi-phase ISM and trace star formation throughout a galaxy.  There have been several multi-wavelength surveys in recent years \citep[e.g.][]{R07,bigiel,paws} which have mapped the gas, dust, and stellar distribution of several nearby galaxies, including M31 and M33.  Multi-wavelength studies can identify the relationship between stars and the surrounding molecular gas, and reveal whether the relations present at small scales locally, such as the \citet{larson81} scaling relations often used to describe local clouds, also exist on larger scales.  

The empirical scaling relations of \citet{larson81} are considered to be integral to our understanding of the fundamental nature of molecular clouds as they link three key properties which define cloud structure: the mass, size, and velocity dispersion.  However, in cases where observations achieved great sensitivity and high resolution, such as the study by \citet{heyer2009}, molecular clouds were found to span a wide range of surface densities, in contrast with the near constant surface densities measured by \citet{larson81}.  \citet{lombardi2010} concluded that the surface density for molecular clouds required to satisfy Larson's scaling relations is only constant for a given detection threshold, and the value of the surface density is a function of that threshold.  

Due to the varying surface density of molecular clouds, several authors \citep[e.g.][]{paper1, heyer2009} are now examining their clouds with respect to the relation between the size-linewidth coefficient, $v_o$ = $\sigma$/$R^{1/2}$, and the surface density, $\Sigma$.  The $v_o$-$\Sigma$ scaling relation delineates populations of clouds which are bound, unbound, or in virial or pressure-bounded equilibrium \citep{balle2011, field2011, heyer2009}.  This relation allows us to better understand the internal structure and dynamics of molecular clouds as well as the nature of the ISM environment in our own Galaxy.

The $v_o$-$\Sigma$ relation can also guide our understanding of clouds in an extragalactic context, such as in the examination of the relation on global scales in a recent study by \citet{leroy2015} who use the relation to compare populations of clouds in different environments for several nearby galaxies.  The authors find that many of the GMCs in the Milky Way and in nearby galaxies that satisfy the scaling relation are in virial equilibrium without external pressure confinement; however, there is significant scatter in the data indicating that some of the clouds in these galaxies could be marginally unbound.  The clouds which are highly unbound were those in the Galactic centre and in the outer disc of the Milky Way, which suggests that environment has an effect on the virial state of molecular clouds.  The effect of the environment on the structure and evolution of molecular clouds will be explored throughout this work.

In this paper, we explore the properties of clouds formed in a simulated galactic disc.  We investigate how the formation and evolution of these clouds and their stars are affected by their environment.  Using clouds identified from synthetic column density maps, we compare the Larson and $v_o$-$\Sigma$ scaling relations and distributions of properties for our cloud populations to observations of molecular clouds in nearby galaxies.  By studying the simulated cloud populations with respect to their location in the disc, we can determine whether the environment, particularly the galactic shear rate, affects the global properties of molecular clouds and star formation.  

In section~\ref{sec:methods4}, we describe the details of the simulation and the methods by which the synthetic observations were produced and the clouds were identified.  In section~\ref{sec:results4}, we present our results and conclude with a summary and discussion on the implications of our findings in section~\ref{sec:conclusions4}.

\section{Methods}\label{sec:methods4}

Using the smoothed particle hydrodynamics (SPH) code \textsc{Gasoline} \citep{gasoline}, we simulated a large-scale galactic disc.  The model we adopt is similar to the models of \citet{tasker2009}, \citet{tasker2011}, and \citet{dobbs11b}, and includes a static galactic potential, self-gravity, mechanisms for heating and cooling, star formation, and stellar feedback.  The disc properties are summarized in Table~\ref{table:discprops}.  The total gas mass in the simulation is of the same order as that in the Milky Way and the spiral galaxy M33 \citep{corbelli}.  The overall global properties of our disc, including the molecular gas surface density (when $n$ $>$ 100 cm$^{-3}$) and galactic shear, are also well-matched to those for M33 \citep{corbelli,heyer2004}.  We used the same logarithmic dark matter halo potential as \citet{tasker2009} and \cite{dobbs11b} to produce an approximately flat rotation curve, typical of observed spiral galaxies and described by the rotational velocity of the galactic disc,
\begin{equation}
	v = \Omega R_{\text{gal}} = \frac{v_{\text{c}} R_{\text{gal}}}{\sqrt{R_{\text{0}}^{\,2} +R_{\text{gal}}^{\,2}}}, \label{eq:halopot}  
\end{equation}
\citep{BT08} where $v_{\text{c}}$ = 220 km s$^{-1}$ is the constant rotational velocity at large radii chosen to be comparable to the Milky Way rotational velocity at the solar radius (8 kpc).  When $R_{\text{gal}}$  $\gg$  $R_{\text{0}}$ = 1 kpc, the rotational velocity is constant at $v$ = $v_{\text{c}}$.  The disc was initialised with a Kolmogorov turbulence spectrum, producing random velocity perturbations in the x, y, and z directions.  The resulting surface density profile of the disc is truncated between 12 and 14 kpc and normalized to 14 M$_{\odot}$ pc$^{-2}$ at the solar radius.  We also excluded the dense, compact centre of the galaxy ($R_{\text{gal}}$ $<$ $R_{\text{0}}$) to conserve resolution in our simulation.  We use the stochastic star formation criteria outlined in \citet{stinson} to determine whether a gas particle of mass $m_{\text{gas}}$ is eligible to be converted into a star particle, which in our case represents a cluster of stars with mass $m_{\text{star}}$ determined by a core-to-star efficiency factor of 30\% \citep[e.g.][]{ladalada}.  To determine the probability that a star will form, we use a constant galactic star formation efficiency per dynamical time, $c_*$, of 6\%, which is typical for cosmological and galactic disc simulations \citep[e.g.][]{stinson, fabio, TB08}.  We adopt a stellar feedback prescription based on the model proposed by \citet{agertz}, where supernovae feedback energy is deposited to the ISM via the nearest gas particle to the star and is converted from a non-cooling state to a radiative cooling form of energy over a timescale of 5 Myr \cite[see also][Benincasa et al., in preparation]{keller2014}. We explore a case where 10\% of the total feedback energy is available to be delivered throughout the disc, corresponding to 10$^{50}$ erg of energy.  

\begin{table}
	\caption{Properties of the global galactic disc simulation}
	\centering
	\begin{tabular}{c c}
	\\
	\hline\hline   
	Gas mass & 7.5 $\times$ 10$^{9}$ M$_{\odot}$ \\
	Radius & 15 kpc \\
	Mass resolution, $m_{\text{gas}}$ & 441 M$_{\odot}$ \\
	$\Sigma$(r = 8 kpc)	   &	14 M$_{\odot}$ pc$^{-2}$  \\
 	Scale height	&	300 pc \\
	Star formation efficiency, $c_*$	&	6 \% \\
	Feedback efficiency & 	10 \% \\
	\hline
  \end{tabular}
  \label{table:discprops}
\end{table}

We used synthetic column density maps of the entire galactic disc to compare our results more directly to observations.  To create these maps, we interpolated the SPH particles onto a 5000 $\times$ 5000 grid using a kernel function \citep{ML85} similar to the method of \citet{ward2012}.  Each map has a grid spacing of 6 pc, which was chosen to match the 1.$\arcsec$5 resolution of the M33 observations made by \citet{R03}.  The gravitational softening length is 20 pc with a minimum smoothing length of 10 pc.  By smoothing our column density map using a Gaussian, we can match the effective resolution of observations by selecting a full-width half-maximum (FWHM) equal to the beam width of a given telescope while also ensuring that our resolution is greater than our softening.  As a result, our surface density maps have an effective spatial resolution of 24 pc.

\begin{figure*}
	\begin{center}
	\includegraphics[scale=1.4,clip=true,trim=4cm 1cm 4cm 0cm]{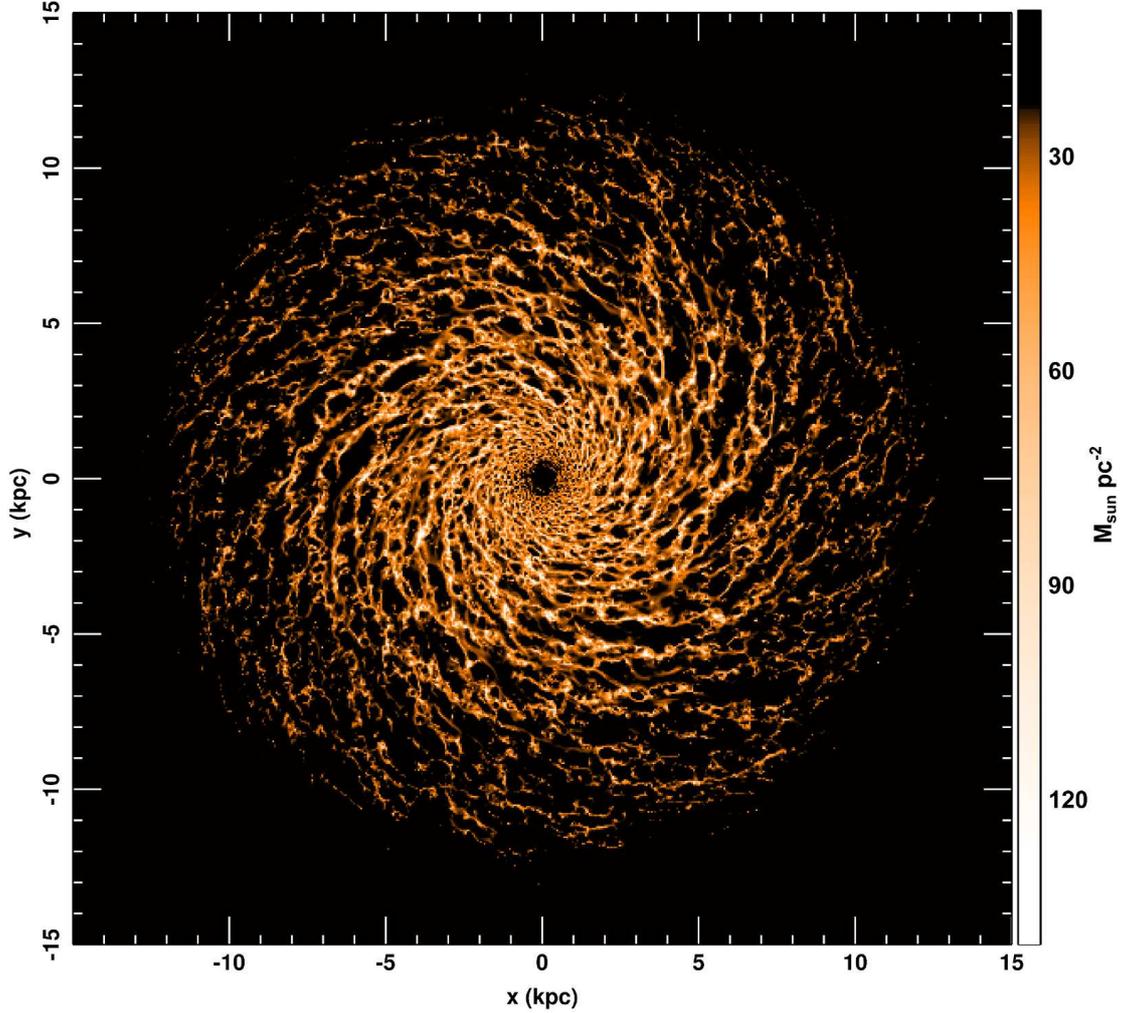}
	\caption[Synthetic column density map of a galactic disc at 500 Myr]{\label{fg:maps} -- Synthetic column density map of a galactic disc at 500 Myr.  The map shows the gas with surface density above a threshold of 10 M$_{\odot}$ pc$^{-2}$.} 
	\end{center}
\end{figure*}

In our analysis, the individual SPH particles must satisfy the condition that their densities are greater than 100 cm$^{-3}$, corresponding to the threshold density to form molecular hydrogen.  The synthetic column density map is shown in Figure~\ref{fg:maps}.  

We identified clouds in our disc using the two-dimensional \textsc{cupid} implementation of the source extraction algorithm, Clumpfind \citep{clumpfind}, in the \textsc{Starlink} software environment\footnote{http://starlink.eao.hawaii.edu}.  The clouds are selected using a surface density threshold of $\sim$ 10 M$_{\odot}$ pc$^{-2}$ and require a minimum number of 7 pixels.  This surface density threshold is estimated from the off-source background and is approximately twice the value of the equivalent rms surface density, $\sigma_{\text{rms}}$, used by \citet{R03}.  Each contour interval was equal to approximately 3$\sigma_{\text{rms}}$.  Clouds that did not extend above the second contour level and clouds with sizes smaller than the effective beam width were rejected.  To account for the added smoothing, the deconvolved cloud sizes were recorded.  Also, since our mass resolution is $\sim$ 440 M$_{\odot}$ per particle, we define our completeness limit at 4.4 $\times$ 10$^4$ M$_{\odot}$, as clouds with masses below this limit will be resolved by less than 100 particles.  

\section{Results}\label{sec:results4}

We examined the properties of molecular clouds identified in our disc at early (250 Myr) and late (500 Myr) times.  These times represent approximately one and two rotations of the outer disc, respectively.  From the maps, we measured the area, $A$, and gas mass, $M$, of each cloud above the surface density threshold.  We then calculated the radius of each cloud assuming spherical geometry such that $R = (A/\pi)^{1/2}$.  The velocity dispersions determined for each cloud were calculated based on the mass-weighted velocities of the particles assigned to that cloud.  Particles with $n$ $>$ 100 cm$^{-3}$ were assigned to clouds if their line-of-sight positions intersected with pixels belonging to a given cloud.  We used one-dimensional velocity dispersions, $\sigma_{\text{v}}$, for our clouds to compare with the line-of-sight velocity dispersions measured in observations.  Using the observable properties of mass, radius, and velocity dispersion, we calculated the virial parameter, 
\begin{equation}
	\alpha = \frac{5\sigma_{\text{v}}^2 R}{G M},
\end{equation}
\citep{BM92} which describes simple virial equilibrium where $\alpha$ = 1 for virialised clouds and clouds with $\alpha$ $\leq$ 2 are gravitationally bound.  We also measured properties of the stars, in order to study the star formation with respect to the properties of the clouds in which they form.  To determine the star formation rate (SFR) for each cloud, we assigned each star particle to the cloud they coincide with along the line-of-sight.  We then calculated the total mass of the stars per cloud and divided by the output time.  The star formation rate surface density, $\Sigma_{\text{SFR}}$, is calculated by dividing the SFR by the cloud area.  These calculations were done at both output times.  The median values and median absolute deviations for these properties are summarized in Table~\ref{table:cloudprops}.

\begin{table*}
	\caption{Median Properties for Molecular Clouds}   \label{table:cloudprops}
	\begin{center}
	\begin{tabular}{c c c}
	\\
	\hline\hline
	& $t$ = 250 Myr & $t$ = 500 Myr \\
	\hline
	$N_{\text{clouds}}$ & 1398 & 1557 \\
	$M$ (10$^{5}$ M$_{\odot}$) & 1.4 $\pm$ 0.6 & 1.6 $\pm$ 0.7\\
	$R$ (pc) & 36 $\pm$ 5 & 36 $\pm$ 5\\
	$\sigma_{\text{v}}$ (km s$^{-1}$) & 2.4 $\pm$ 0.6 & 2.5 $\pm$ 0.6 \\
	$v_o$ = $\sigma_{\text{v}}$/$R^{1/2}$ (km s$^{-1}$ pc$^{-1/2}$) & 0.4 $\pm$ 0.1 & 0.4 $\pm$ 0.1\\
	$\alpha$ & 1.5 $\pm$ 0.5 & 1.5 $\pm$ 0.5\\
	$\Sigma$ (M$_{\odot}$ pc$^{-2}$) &  35 $\pm$ 9 & 40 $\pm$ 10 \\
	$\Sigma_{\text{SFR}}$ (10$^{-3}$ M$_{\odot}$ yr$^{-1}$ kpc$^{-2}$)& 6 $\pm$ 3 & 5 $\pm$ 2  \\
	\hline
  \end{tabular}
  \end{center}
  \footnotesize{\textbf{Notes:} $N_{\text{clouds}}$ is the total number of clouds, $M$ is the cloud mass, $R$ is the cloud radius, $\sigma_{\text{v}}$ is the one-dimensional velocity dispersion where 3$\sigma_{\text{v}}^2$ = ($\sigma_{\text{v}_{\text{x}}}^2$+$\sigma_{\text{v}_{\text{y}}}^2$+$\sigma_{\text{v}_{\text{z}}}^2$), $v_o$ is the size-linewidth coefficient, $\alpha$ is the virial parameter where $\alpha$ = 2 for clouds in gravitational equilibrium, $\Sigma$ is the gas surface density, and $\Sigma_{\text{SFR}}$ is the star formation rate surface density.  The median and median absolute deviation is listed for all cloud properties for direct comparison to the results for M33 provided in \citet{hughes13}.}
\end{table*}
  
We find that the number of clouds increased with time between 250 Myr and 500 Myr and that $\sim$ 30\% of clouds were unbound, demonstrating that a substantial fraction of unbound clouds form naturally in our disc \cite[see also][]{dobbs11,dobbs11b,tasker2009}.  We also find that the resultant star formation rate surface densities ($\Sigma_{\text{SFR}}$) of clouds in this disc are close to -- and slightly lower than -- those expected from the Kennicutt-Schmidt relation, which shows that the presence of unbound clouds in our disc helps to match the low galactic star formation rates observed in the Milky Way \citep{robitaille2010} and nearby galaxies \citep{kennicutt,bigiel}. 

To understand these results in the context of the formation and interactions of clouds in the disc, we calculated the angular momentum vector for each cloud and determined the angle offset, $\theta$, from the angular momentum vector of the galactic disc.  This angle represents the axial tilt of the cloud with respect to the disc.  A cloud with $\left|\theta\right|$ $<$ 90\textdegree\, is rotating the same direction as the disc, or prograde, and those with 90\textdegree $<$ $\left|\theta\right|$ $<$ 180\textdegree\, are retrograde with respect to the disc.  Clouds which form in the disc due to a gravitational instability will inherit the direction of the disc's rotation, resulting in a cloud that is rotating prograde relative to the disc.  Therefore, if clouds are found to be rotating retrograde with respect to the disc, we can assume that they have been perturbed by an interaction with the disc or by a collision with another cloud \citep{tasker2009}.  We find that the overwhelming majority of clouds are prograde as expected for clouds forming in a disc environment; however, we do find a significant population (13\%) of clouds with retrograde rotation at 250 Myr.  The percentage of clouds with retrograde rotations does not increase with time, which suggests that the low amount of energy injection from feedback results in few cloud collisions and interactions over time and more star formation.  Since there was little disruption of the disc due to the low feedback energy from stars, clouds were able to form, grow, and evolve.  

\subsection{Comparison to observations}

\begin{figure*}
	\begin{center}
	\includegraphics[scale=0.77, clip=true, trim=1.5cm 0 0 0]{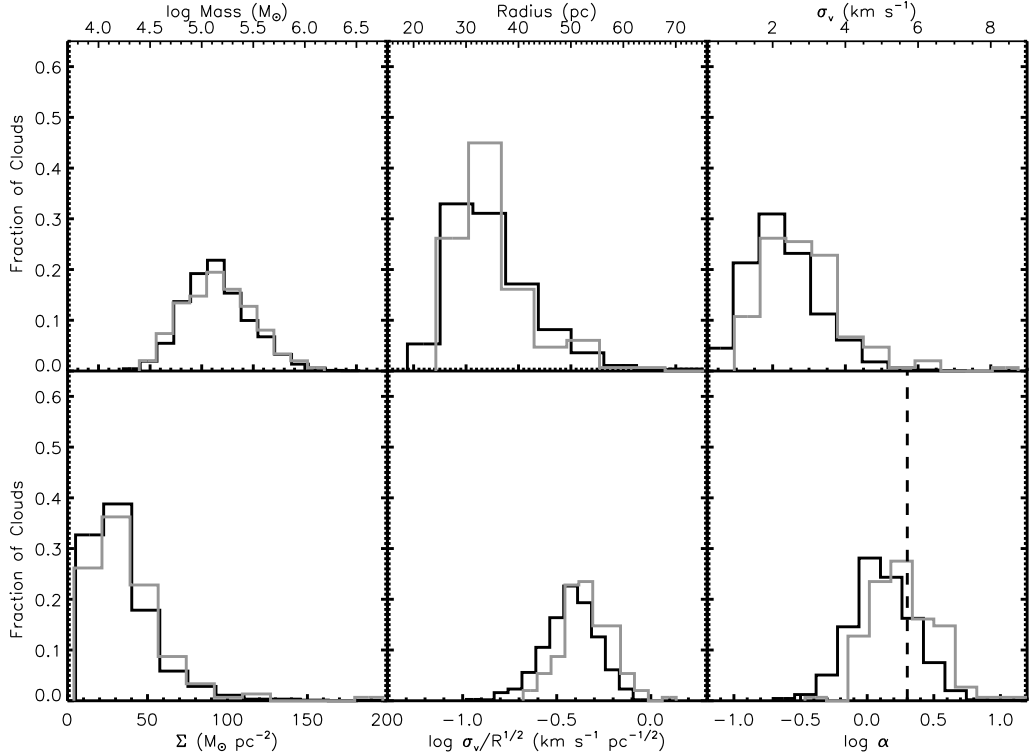}
	\caption[Distributions for the mass, radius, velocity dispersion, surface density, size-linewidth coefficient, and virial parameter of molecular clouds in our sample compared to those for M33]{\label{fg:m33histograms} -- Distributions for the mass, radius, velocity dispersion ($\sigma_{\text{v}}$), surface density ($\Sigma$), size-linewidth coefficient ($v_o$), and virial parameter ($\alpha$) of molecular clouds in our sample (black histograms) with 10\% feedback efficiency at 500 Myr.  The corresponding distributions for the clouds identified in M33 \citep{R07} are shown as grey histograms for comparison.  The vertical dashed line in the bottom-right panel represents the boundary to the left of which clouds are gravitationally bound. }
	\end{center}
\end{figure*}

At a distance of 0.84 Mpc \citep{FWM91}, M33 is one of the nearest galaxies to the Milky Way which makes it a prime candidate for comparison to our simulations as, not only are its gross properties comparable to the simulation, but its clouds are observationally well resolved at this distance.  We identified 1557 molecular clouds in the simulated galactic disc at 500 Myr.  In Figure~\ref{fg:m33histograms}, we compare the distributions of cloud properties for the simulated clouds (black histograms) to those for 149 real molecular clouds (grey histograms) from multi-wavelength observations of M33 \citep{R07}. Although the grid spacing and resolution of our synthetic column density maps match the M33 observations of \citet{R03}, the less-resolved sample of observed clouds by \citet{R07} is more appropriate for comparison as it uses a merged multi-wavelength data set to produce a fully-sampled map of the entire galaxy.   The authors find that the majority of the clouds detected lie within a galactocentric radius of 4 kpc, resulting in a number density of approximately 3 clouds kpc$^{-2}$, consistent with the number density of clouds in our comparatively larger simulated disc.  We find excellent agreement between the observed and simulated cloud distributions for the masses, sizes, and surface densities, as confirmed by a two-sided Kolmogorov-Smirnov (KS) test.  The KS test showed statistical differences between the observed and simulated cloud distributions for velocity dispersion and its related properties, $v_o$ and $\alpha$; however, we see that the medians and ranges of values for these properties in our simulated clouds are similar to those in the observed clouds.  These results demonstrate that the global properties of our simulated clouds are comparable to those in M33.

\begin{figure*}
	\centering
	\subfigure{\includegraphics[scale=0.58,clip=true, trim=5mm 0 0 0]{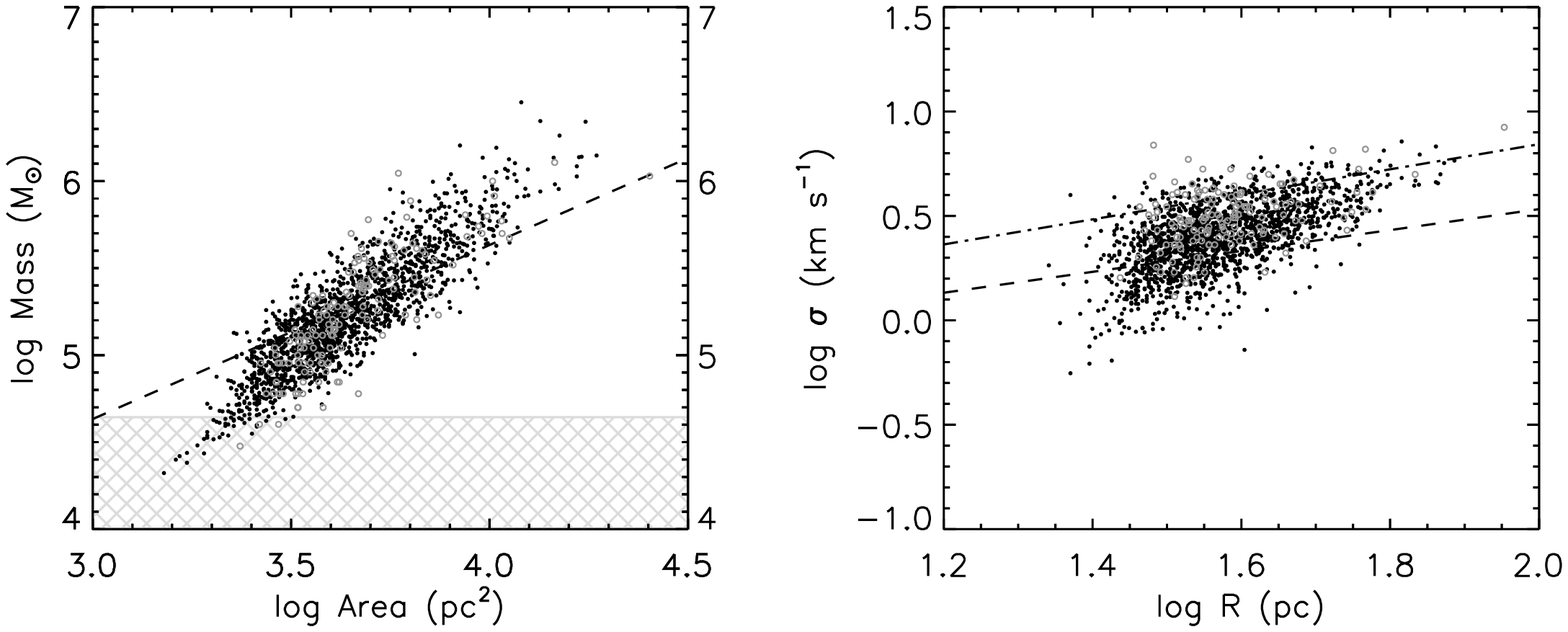}}	
	\subfigure{\includegraphics[scale=0.57,clip=true, trim=5mm 0 0 0]{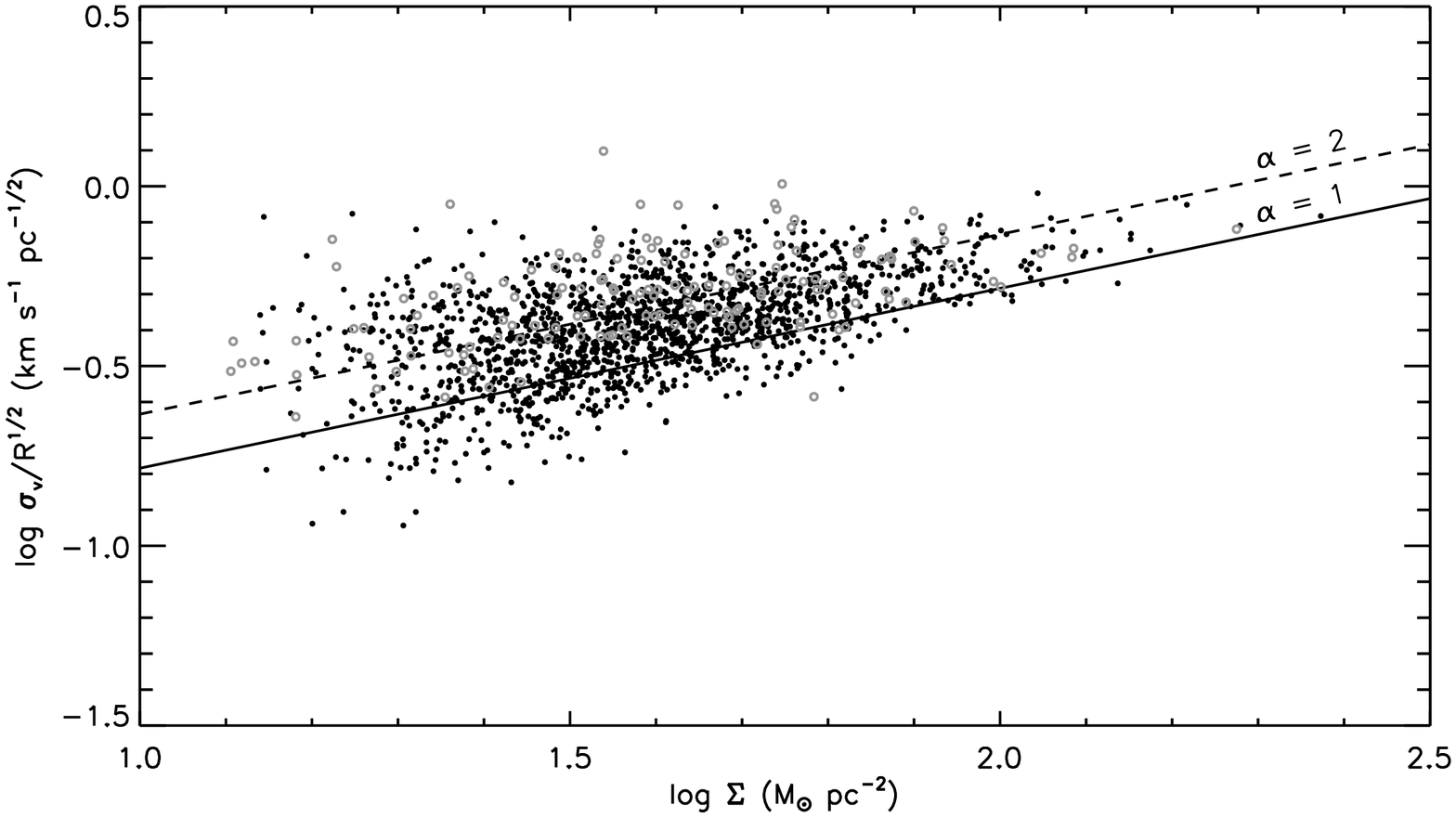}}	
	\caption[Scaling relations for molecular clouds extracted from synthetic column density maps]{\label{fg:scaling} -- Scaling relations for molecular clouds extracted from synthetic column density maps.  Mass as a function of cloud area (top left) and linewidth as a function of cloud radius (top right) are shown with their corresponding empirical scaling relation of \citet{larson81} (dashed black lines).  Molecular cloud data from observations of M33 by \citet{R07} are also shown (grey circles).  The grey hashed region in the top left panel shows masses for clouds with less than 100 particles.  The dot-dashed line in the top right panel shows the scaling relation derived from observations of extragalactic clouds by \citet{bolatto} for comparison.  The size-linewidth coefficient, $v_o$, as a function of the surface density (bottom) is also shown, where the dashed and solid black lines represent the boundaries below which clouds are gravitationally bound ($\alpha$ = 2) and virialised ($\alpha$ = 1) respectively.}
\end{figure*}

We also find that our clouds lie on the scaling relations of \citet{larson81} determined from Galactic molecular cloud data, shown by the dashed black lines in the top left and right panels of Figure~\ref{fg:scaling}.  The scaling relation derived from observations of extragalactic molecular clouds by \citet{bolatto} is also shown as a dot-dashed line in the top right panel for comparison to our results.  We find that, although Larson's laws originated from observations of local clouds, the scaling relations are still present on larger scales in our simulations of an extragalactic cloud population.  We note however that there is still a lot of scatter in the observed and simulated data.  In \citet{paper1}, we showed that the scatter accompanying the Larson scaling relations for local clouds was attributed to a spread in the cloud surface density, $\Sigma$, and the intrinsic virial parameter.  We plot the size-linewidth coefficient, $v_o$, as a function of $\Sigma$ to determine the spread in the virial parameter at fixed surface density and we see that both the simulated (black points) and observed (grey circles) populations consist of bound and unbound clouds, where gravitationally bound clouds lie below the dashed line where $\alpha$ = 2.  Based on the results of \citet{paper1}, this is not unexpected as we previously showed that a substantial population of unbound clouds can match the observed scaling relations with their accompanying scatter in addition to forming stars; however, we have now demonstrated that unbound clouds can not only match these observations but they form naturally in a galactic disc environment, under the effects of galactic potential, shear, external pressure, and stellar feedback.

Unbound clouds have been simulated in detail in several recent studies in both isolation \citep[e.g.][]{clark05,clark08,bonnell2011,paper1} and in galactic discs \citep[e.g.][]{dobbs11,dobbs11b,HQM12}.  These studies have shown that there are a number of factors, both internal and external, that can affect the virial parameter of a molecular cloud.  High mass clouds tend to be more strongly bound than less massive clouds as seen in the observations of \citet{hcs01} and the models of \citet{HQM12}.  Although several authors argue that these low mass clouds must be in pressure-confined virial equilibrium, the necessary pressures to maintain equilibrium for these broadened linewidths can be upwards of 10$^7$ cm$^{-3}$ K \citep{field2011} for some clouds, much higher than the typical ambient ISM pressure of $P/k_{\text{B}}$ $\sim$ 10$^{3-4}$ cm$^{-3}$ K \citep{Wolfire03}.  An alternative theory is that some of these clouds are simply unbound and short-lived with lifetimes of $\sim$ 10 - 15 Myr \citep[][and references therein]{clark05}.  Molecular clouds can be disrupted by cloud-cloud collisions, galactic shear, and supernovae feedback, all of which are sources of turbulence in the disc.  The more turbulent the molecular gas, the `puffier' the cloud will be for a fixed virial parameter and surface density, as seen in the size-linewidth scaling relation.  As noted earlier in this section, turbulence from supernovae feedback and cloud-cloud collisions likely plays an important role in affecting the boundedness and evolution of molecular clouds.

\subsection{The effect of galactic shear on molecular cloud properties}

\begin{figure*}
	\centering
	\subfigure[250 Myr]{\includegraphics[scale=0.77, clip=true, trim=1.5cm 0 0 0]{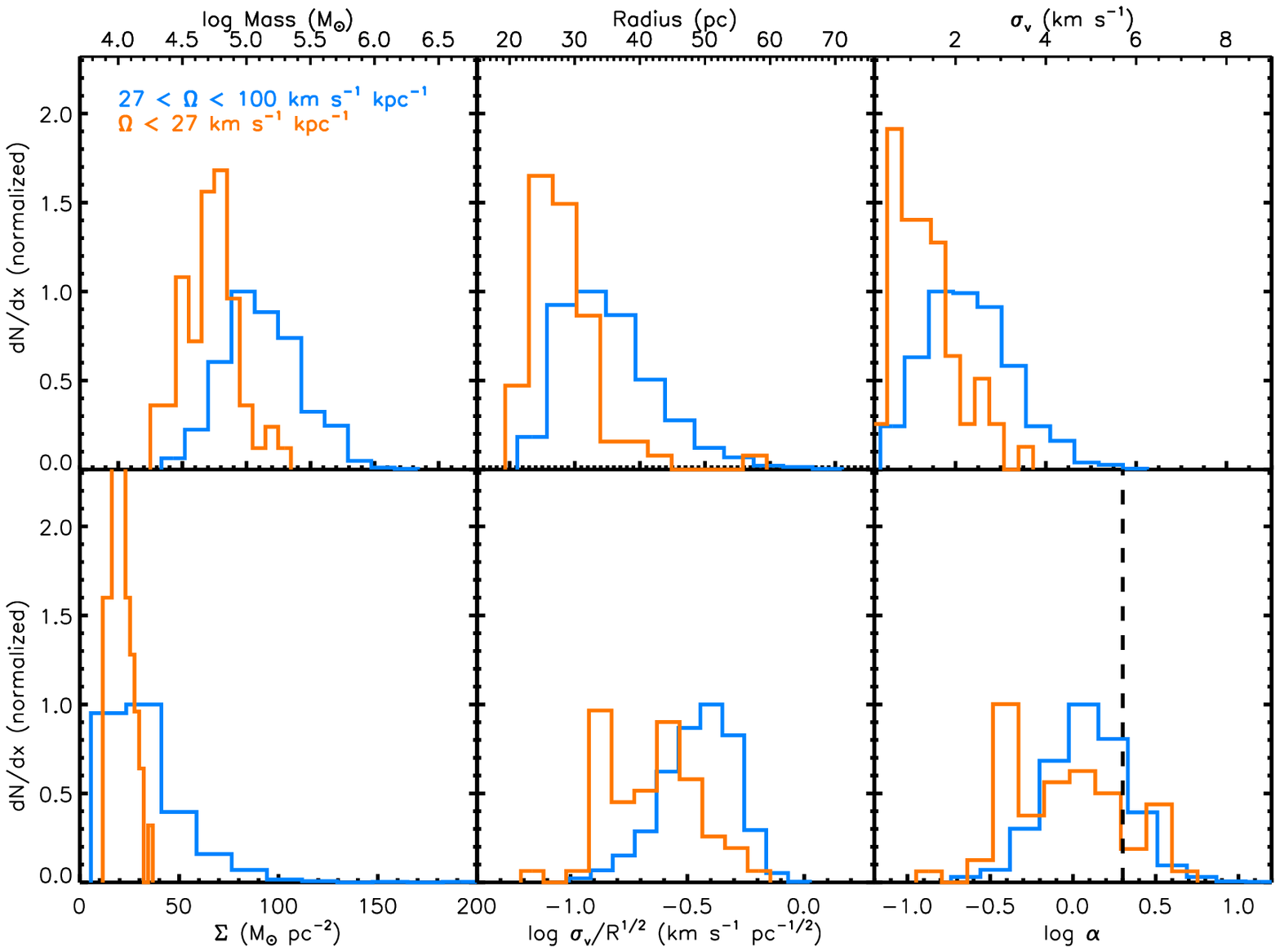}}
	\subfigure[500 Myr]{\includegraphics[scale=0.77, clip=true, trim=1.5cm 0 0 0]{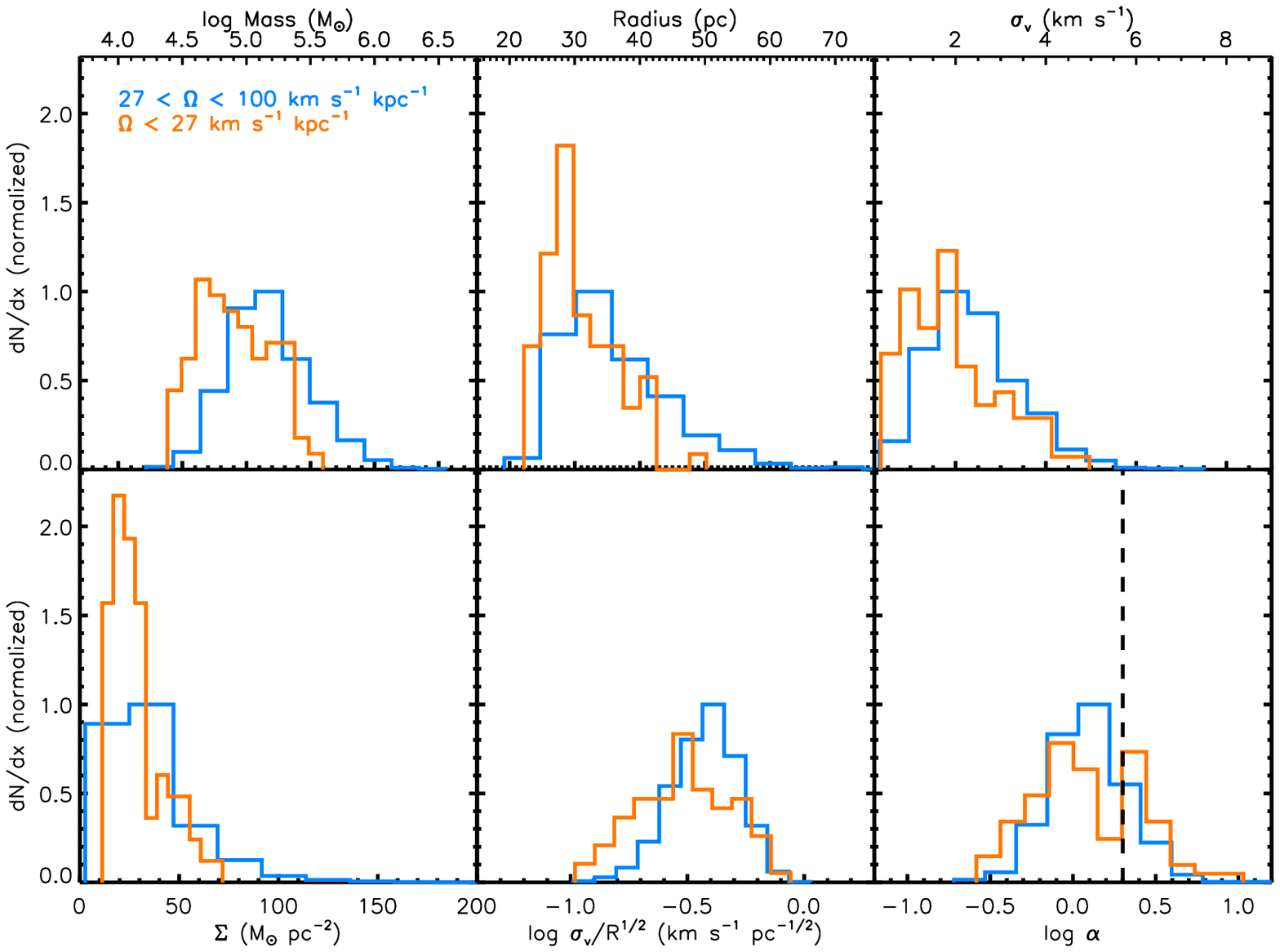}}
	\caption[Cloud properties divided by the differential rotation rate ($\Omega$)]{\label{fg:shearhistograms} -- Cloud properties as described in Figure~\ref{fg:m33histograms} but here the clouds are divided into radial annuli at $R_{\text{gal}}$ =  8 kpc or equivalently by the differential rotation rate ($\Omega$ = 27 km s$^{-1}$ kpc$^{-1}$) for each annulus. }
\end{figure*}

In addition to turbulence driven by feedback energy and cloud-cloud collisions, turbulence can also arise from shear in the galactic disc.  To determine whether the galactic shear rate, $\Omega$, has a significant effect on the properties of molecular clouds, we divided the cloud population into two regions: the inner galaxy, where the shear rate is high, and the outer galaxy, where the shear rate is much lower.  Our choice of the boundary was guided by observations of M33 \citep{bigiel10,R07}, where the outer disc of M33 is defined as the region beyond two CO scale lengths ($R_{\text{gal}}$ $\approx$ 4 kpc).  At this radius, the emission has fallen by a factor of $e^2$.  Since our disc is significantly larger than M33, we chose a division based on a matching shear rate rather than galactocentric radius.  Using the rotation curve of M33 \citep{corbelli}, we determined that its shear rate at 4 kpc is approximately 27 km s$^{-1}$ kpc$^{-1}$.  An equivalent shear rate in our disc is at a galactocentric radius of approximately 8 kpc, determined using equation~\ref{eq:halopot}.  We sorted each cloud by its location in the disc and produced distributions similar to those shown in Figure~\ref{fg:m33histograms} for the inner (2 -- 8 kpc, blue) and outer disc ($>$ 8 kpc, orange), excluding the innermost portion of our disc to avoid any edge effects from the hole in the centre of the model.  The distributions were normalized by the number of clouds per distribution and the bin size and are expressed in units of the peak of the inner disc distribution (blue).  The resulting distributions are shown in Figure~\ref{fg:shearhistograms} for two output times, 250 and 500 Myr. 
 
Figure~\ref{fg:shearhistograms} shows that the distributions differ at early and late times for each cloud property.  We find that the distributions of properties for clouds in the outer disc with $\Omega$ $<$ 27 km s$^{-1}$ kpc$^{-1}$ are shifted to lower masses, sizes, and velocity dispersions.  To measure whether there is a statistically significant difference between the distributions of our inner and outer disc cloud properties, we perform a two-sided Kolmogorov-Smirnov (KS) test on the distributions.   The probabilities that the inner and outer disc clouds are drawn from the same distribution are given in Table~\ref{table:KSclouds}.
 
\begin{table}
	\caption{KS test probabilities for inner and outer disc molecular clouds}   \label{table:KSclouds}
	\begin{center}
	\begin{tabular}{c c c}
	\\
	\hline\hline
	& $t$ = 250 Myr & $t$ = 500 Myr \\
	\hline
	$M$  & 3.95$\times$10$^{-24}$ & 4.13$\times$10$^{-12}$  \\   	
	$R$ & 4.92$\times$10$^{-13}$ & 1.18$\times$10$^{-10}$\\         	
	$\sigma_{\text{v}}$ &  9.27$\times$10$^{-15}$ & 7.89$\times$10$^{-8}$  \\         			
	$\Sigma$  & 5.64$\times$10$^{-24}$ & 5.97$\times$10$^{-12}$ \\	
	$v_o$ = $\sigma$/$R^{1/2}$ & 5.84$\times$10$^{-13}$ & 5.98$\times$10$^{-6}$  \\	
	$\alpha$ & 3.13$\times$10$^{-6}$ & 0.05 \\			
	\hline
  \end{tabular}
  \end{center}
\end{table}

We consider any probabilities, $Pr$, less than 0.05 an indication that there are statistically significant differences between the distributions.  We find that every pair of distributions are statistically different, except for $\alpha$ at 500 Myr which has $Pr$ = 0.05.  To test the robustness of these results, we permuted the order of the clouds using a pseudorandom number generator and split the data into two distributions in four different ways: 25/75, 50/50, 75/25, and by the fractional amount resulting from a division of the sorted data at a boundary of 8 kpc (95/5).   We performed two-sided KS tests on these four pairs of distributions for all properties at both 250 and 500 Myr.  We repeated this process 100 times and calculated the mean probabilities that the paired data could be drawn from the same distribution for each cloud property.  We found that in every case, regardless of the division of data, all properties had $Pr$ $\gtrsim$ 0.5, indicating that it is quite likely that the clouds are similar to one another overall but exhibit distinct statistical differences when they are compared in the context of their environment.

Observations also show that clouds in the outer disc of M33 \citep{bigiel10} and the Milky Way \citep{hcs01,BW95} exhibit lower masses, sizes, and velocity dispersions than those found in the inner disc.  These findings are sometimes attributed to low resolution, blending, incomplete sampling, and, for the case of the Milky Way, line-of-sight confusion in the determination of cloud properties in the inner disc.  However, we find a real difference between the inner and outer disc cloud populations.  

To determine whether our results are dependent on our choice of the boundary, we repeated our statistical comparison of the cloud distributions for the inner and outer disc using the boundary corresponding to one CO scale length or one e-fold of the disc emission, which for M33 is 2.1 kpc \citep{R07}.  Using the shear rate at 2.1 kpc for M33 \citep{corbelli} and equation~\ref{eq:halopot}, we calculate the equivalent radius in our disc at this shear rate to be approximately 5.5 kpc.  We find similar results as before, although with much less pronounced differences between the inner and outer distributions than those seen for clouds within and beyond the 8 kpc boundary, which indicates that clouds in the farthest reaches of the galaxy are the most strongly affected by environmental factors.  Every property still has statistically significant differences with $Pr \ll$ 0.05 except for $\sigma_{\text{v}}$ and $v_{\text{o}}$, which have KS probabilities slightly larger than 0.05 at late times.  These results show that, regardless of the boundary, the properties which are most similar over the entire disc are the distributions for the size-linewidth coefficient and the virial parameter, demonstrating that, while the clouds in the inner portion of the disc can have significantly different observable properties from those in the outer disc, they will still exhibit scaling relations that are consistent regardless of environment.

\citet{R07} also studied the variation of the cloud mass distributions with galactocentric radius in M33 and found a similar result that the distributions significantly differ for inner and outer disc clouds.  In particular, their observations of M33 showed that there were very few molecular clouds in the outer disc beyond 4 kpc, despite the abundance of molecular gas beyond this radius, and that the few clouds found in the outer disc had much lower masses than those within this radius, consistent with our results.

\subsubsection{Toomre analysis}

Molecular clouds appear to be a product of the environment in which they formed; however, it has yet to be determined which environmental factors play a significant role in distinguishing the properties of molecular clouds between the inner and outer disc.  \citet{R07} suggest that cloud properties at small radii could differ due to strong galactic shear and less-defined spiral structure in the region and that the lack of clouds at large radii could be caused by features of the galactic disc, such as its gravitational stability and the surface density distribution of atomic gas.  The observable properties (mass, size, and velocity dispersion) are much smaller for clouds formed in the outer disc where the external pressure (Benincasa et al., in preparation) is weaker.  This may seem counter-intuitive as one might expect cloud size to increase with decreased external pressure, but it can be understood if we account for the effect of galactic shear by a Toomre analysis of a differentially rotating disc.  

\citet{toomre} considered gravity, thermal pressure, and the Coriolis force caused by rotation the three factors that contribute to the stability of a galactic disc.  The Toomre stability criterion which determines the stability boundary at which these factors are balanced is 
\begin{equation}
	Q = \frac{c_{\text{s}}\kappa}{\pi G \Sigma},
\end{equation}
where $\kappa$ is the epicyclic frequency, $c_{\text{s}}$ is the sound speed, and $\Sigma$ is the surface density of the disc.  If the Toomre $Q$ parameter drops below 1, a differentially rotating disc will be unstable to collapse.  To determine the critical wavelength of a perturbation, $\lambda_{\text{crit}}$, above which all perturbations are unstable, we must begin with the dispersion relation for a uniformly-rotating fluid disc with a non-zero sound speed,
\begin{equation}
	\omega^2 = \kappa^2 - 2\pi G\Sigma|k| + c_{\text{s}}^2 k^2,  \label{eq:toomre}
\end{equation}
\citep{BM92} where $\omega$ is the oscillation frequency and $k$ is the wave number.  By solving for the critical wavenumber, $k_{\text{crit}}$, we can then determine the corresponding wavelength, $\lambda_{\text{crit}}$ = 2$\pi$/$k_{\text{crit}}$, and subsequently the Toomre mass,
\begin{equation}
	M_{\text{T}} = \Sigma \lambda_{\text{crit}}^2.	\label{eq:MT}
\end{equation}

\begin{figure*}
	\begin{center}
	\includegraphics[scale=1]{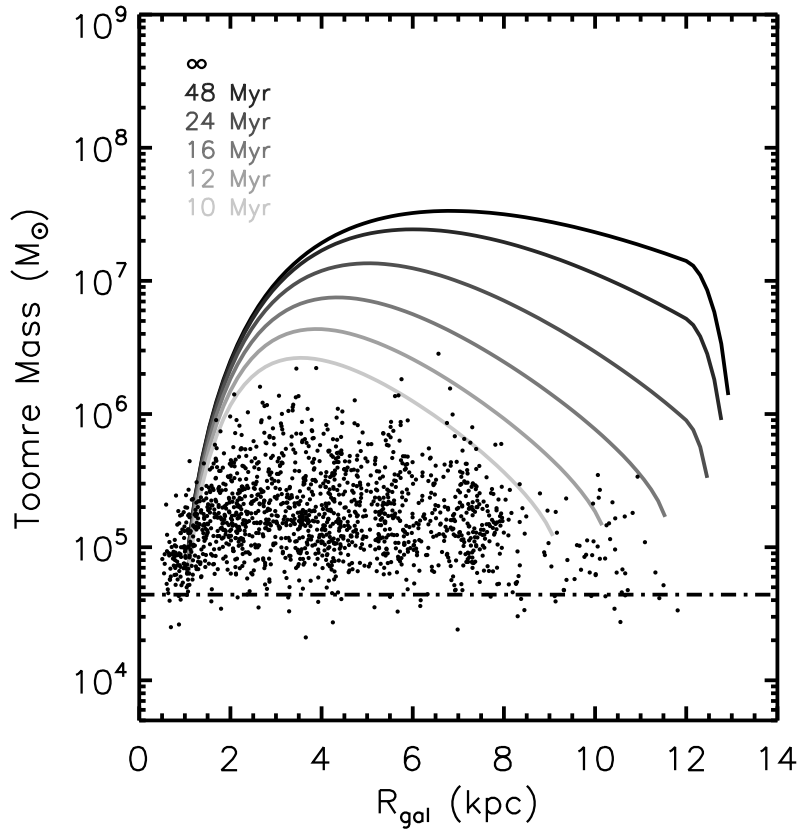}
	\caption[Toomre mass, $M_{\text{T}}$, as a function of galactocentric radius]{\label{fg:toomremass} -- Toomre mass, $M_{\text{T}}$, as a function of galactocentric radius.  The solid lines represent the boundary above which the disc is supported against collapse by shear and are the solutions to equation~\ref{eq:MT} for six collapse timescales, $\tau$ = [$\infty$, 48 Myr, 24 Myr, 16 Myr, 12 Myr, 10 Myr].  The gas masses for clouds identified in our simulated galaxy at 500 Myr are also shown (black dots). The dot-dashed line at 4.4 $\times$ 10$^{4}$ M$_{\odot}$ shows the mass of a cloud resolved by 100 particles.}
	\end{center}
\end{figure*}

The Toomre mass, $M_{\text{T}}$, is analogous to the Jeans mass for the case where the shear rate is an important component to stability.  Therefore, we expect regions of the disc unstable against gravitational collapse to form fragments with masses given by the Toomre mass.  Since the mass is dependent on the surface density of the disc and the critical wavelength, which are both a function of $R_{\text{gal}}$, the mass also scales with the galactocentric radius as shown in Figure~\ref{fg:toomremass}.  The solution for the Toomre mass when $\omega$ = 0, corresponding to an infinite timescale for collapse, is shown in Figure~\ref{fg:toomremass} as a black solid line and represents the line of neutral stability in the disc.  Our simulated disc is not pressure-supported ($dP/d\rho$ $<$ 0) in the regime where gas densities are $\rho$ = 1 - 10 cm$^{-3}$, typical for the ISM.  Therefore, there is a singular solution to equation~\ref{eq:toomre}, representing the boundary above which the disc is stable against collapse by galactic shear.  

To determine whether there is a characteristic timescale for fragmentation in the disc described by $\omega$, we calculated solutions to equation~\ref{eq:MT} for a range of collapse timescales, $\tau$ = 1/$\omega$,  and compared the results to our simulated clouds.  While the Toomre mass remained fairly constant throughout the disc when $\tau\to\infty$, as $\tau$ decreased, the Toomre mass profile began to decrease more with increasing galactocentric radius and therefore we would expect smaller and lower mass clouds to fragment out of the outer disc.  This result supports what we observe in our simulated clouds (shown in Figure~\ref{fg:toomremass} as black dots) as our population of clouds in the outer disc were found to have lower masses and sizes than those in the inner disc.  By confining the cloud data within the envelope of instability created by the solution to equation~\ref{eq:MT}, we find that a collapse timescale between 10 and 16 Myr best describes our data with the solutions shown in Figure~\ref{fg:toomremass} as grey solid lines.  We therefore conclude the characteristic formation timescale for clouds in our disc is $\tau_{\text{form}}$ = 13 $\pm$ 3 Myr.  This collapse timescale is comparable to estimates for the lifetime of a molecular cloud within uncertainties \citep[e.g.][]{murray11,dobbsPPVI}, indicating that the formation and destruction of molecular clouds are in approximate equilibrium with one another.

We find the most significant effect on the properties of the molecular clouds in our disc comes from the galactic shear which causes an overall narrowing envelope of instability at large galactocentric radii, as shown by this Toomre analysis.  As we discovered earlier, the properties of clouds in the outer disc will have lower masses and smaller sizes and this result can be understood as an effect of their environment, due to an increasingly limited range of possible Toomre masses for collapsing gas fragments at large galactocentric radii.
 
\subsection{The effect of environment on star formation rate}
 
\begin{figure*}
	\begin{center}
	\includegraphics[scale=0.8, clip=true, trim=1cm 0 0 0]{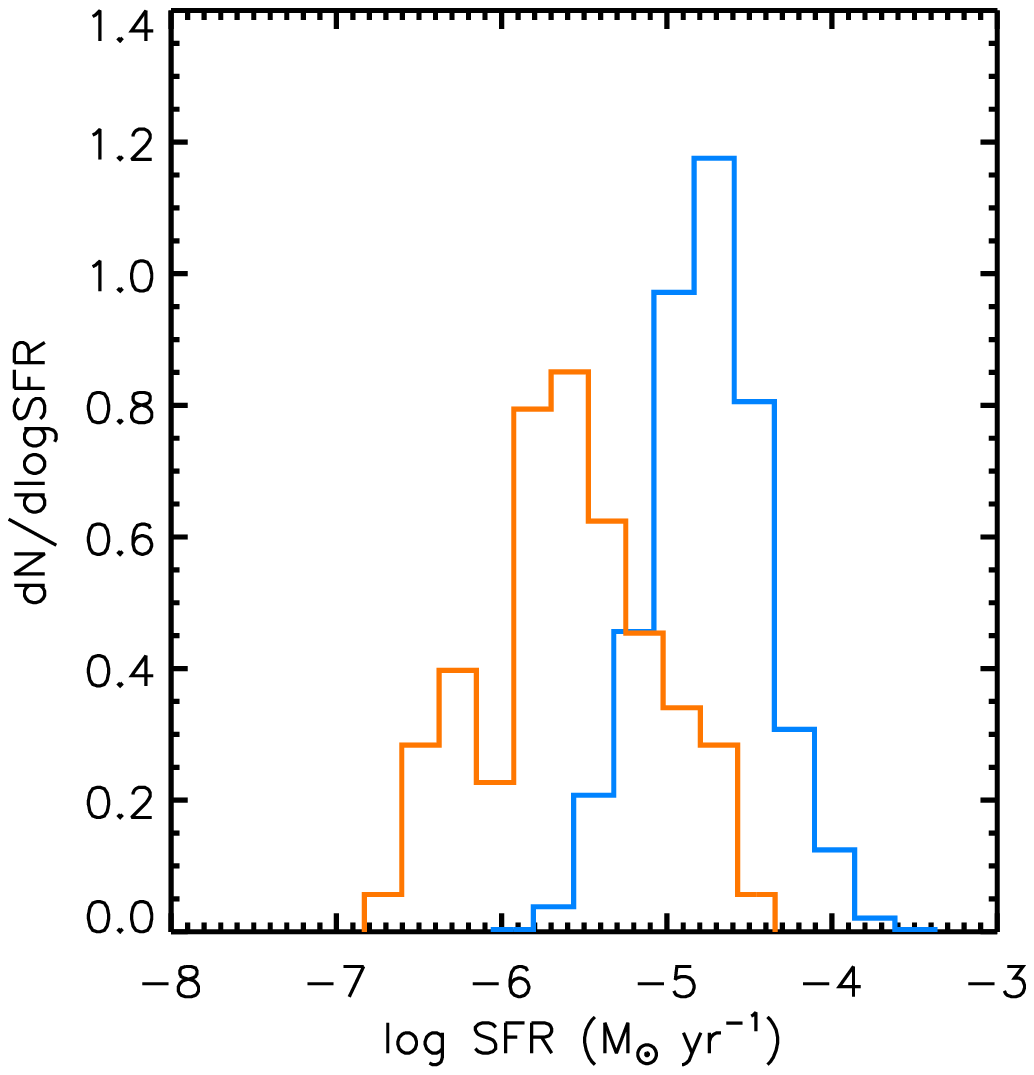}
	\caption[Distributions for the star formation rate (SFR) of our inner disc and outer disc molecular clouds at 500 Myr]{\label{fg:sfr_omega} -- Distributions for the star formation rate (SFR) of our inner disc (blue histogram) and outer disc (orange histogram) molecular clouds at 500 Myr.  The inner and outer disc are divided at 8 kpc (see text).  The distributions are normalized by the number of clouds and by the bin size.} 
	\end{center}
\end{figure*}

We found that the environment, particularly the galactic shear rate, can have a significant effect on the gas properties of molecular clouds, but we have not yet explored the effect, if any, it has on the star formation.  In Figure~\ref{fg:sfr_omega}, we show the SFR distribution for clouds in the inner (2 $< R_{\text{gal}} <$ 8 kpc, blue histogram) and outer disc ($R_{\text{gal}} >$ 8 kpc, orange histogram).

\begin{figure*}
	\begin{center}
	\includegraphics[scale=0.8, clip=true, trim=1cm 0 0 0]{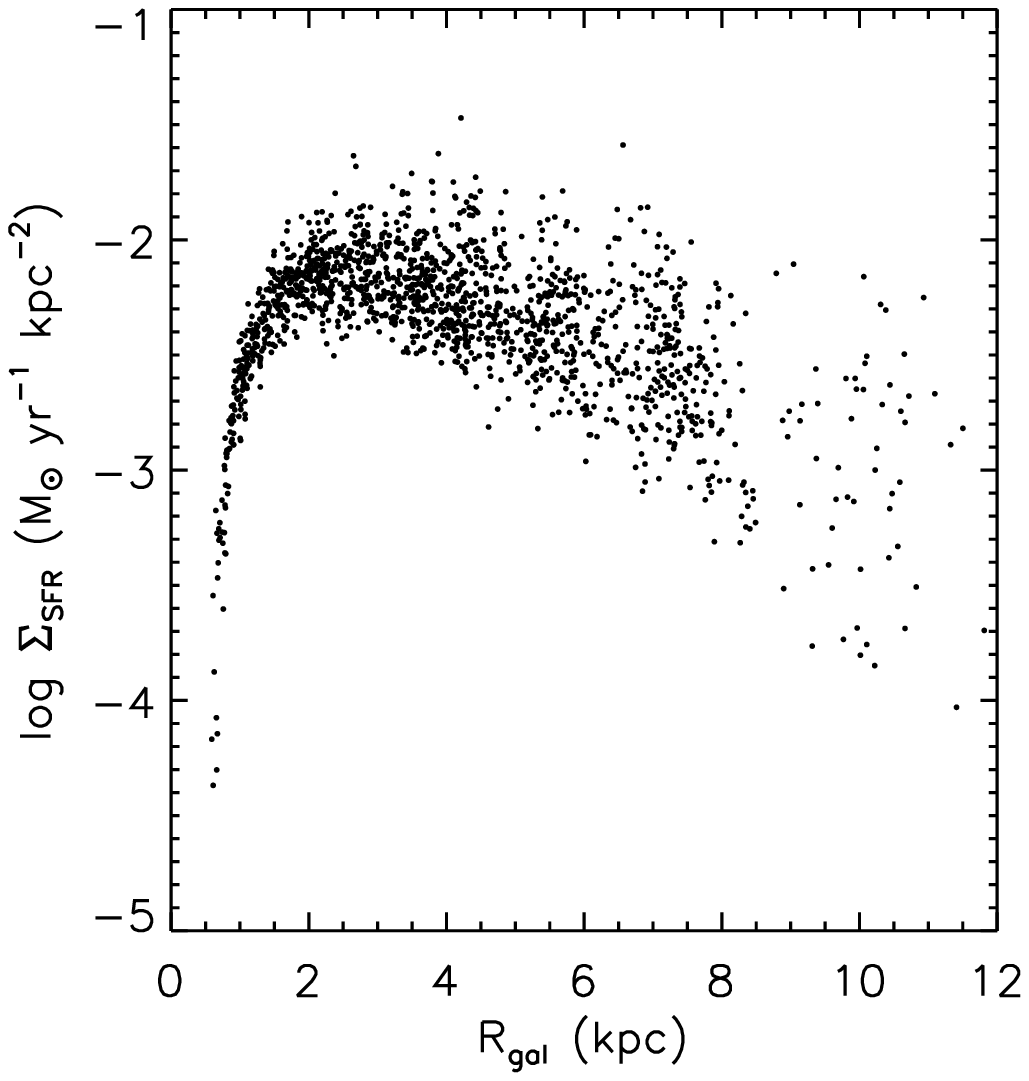}
	\caption[Star formation rate surface density ($\Sigma_{\text{SFR}}$) as a function of the galactocentric radius for molecular clouds in our sample at 500 Myr]{\label{fg:sfrrad} -- Star formation rate surface density ($\Sigma_{\text{SFR}}$) as a function of the galactocentric radius for molecular clouds in our sample at 500 Myr.  Note the similar shape to the shear-supported boundary for the Toomre mass in Figure~\ref{fg:toomremass}. }  
	\end{center}
\end{figure*}

We find that the SFR is significantly higher in inner disc clouds than it is in outer disc clouds, indicating that the star formation is dependent on the location of their parent cloud in the disc.  Figure~\ref{fg:sfrrad} shows the star formation rate surface density ($\Sigma_{\text{SFR}}$) as a function of galactocentric radius in the disc and we find a similar profile to the shear-supported boundary for the Toomre mass in Figure~\ref{fg:toomremass}.

We have already shown that the inner disc is populated with larger, more massive clouds, a product of gravitational instabilities resulting from galactic shear.  We find that the SFR and the corresponding $\Sigma_{\text{SFR}}$ can also be influenced by environmental factors as well, resulting in lower, more inefficient, star formation rates in the outer disc, due to the underlying exponential gas surface density profile of the disc.   This relationship between the SFR and the gas surface density is most commonly described using the Kennicutt-Schmidt star formation law \citep{kennicutt,schmidt}, where 
\begin{equation}
	\Sigma_{\text{SFR}} \propto \Sigma^{1.4}.  
\end{equation}
\citet{bigiel} studied the inner discs of several nearby galaxies and found that $\Sigma_{\text{SFR}}$ was more closely correlated with the molecular H$_2$ surface density than with the HI or total gas surface densities.  However, after studying the outer discs of nearby galaxies and finding a strong trend between the HI surface density and the SFR, \citet{bigiel10a} concluded that the HI surface density profile was a key environmental factor causing inefficient star formation in the outer discs of galaxies.  We find that our simulated disc has less efficient star formation in its outer disc due to the shape of the surface density profile as it decreases with galactocentric radii, consistent with the findings of \citet{bigiel10a}.  

\section{Summary and Discussion}\label{sec:conclusions4}

We simulated a large-scale galactic disc to explore the effects of environment on the structure of molecular clouds.  We found that a substantial population of unbound clouds formed naturally in a galactic disc environment.  We found that the distributions of properties for our clouds were consistent with those measured in observations of M33 by \citet{R07}.  We also showed that our cloud sizes, masses, and velocity dispersions satisfied the Larson scaling relations, confirming the presence of these relations in clouds at extragalactic distances on global size scales.  We conclude that a mixed population of bound and unbound clouds formed in a galactic disc under the influence of turbulence, gravity, galactic potential, shear, stellar feedback, and external pressure can match observations when studied in an extragalactic context.

By separating our clouds into two populations based on their location in the disc, we explored their properties in the context of their environment.  Using Toomre analysis, we found galactic shear to be an important environmental factor affecting the upper limit of cloud properties in the outer disc, particularly due to a narrowing envelope of instability at large radii caused by a decrease in the timescales for collapse.  This narrowing produces a lower Toomre mass for fragmentation than in the inner disc, resulting in smaller, lower mass clouds in the outer disc.  We also explored the effect of environment on stars and found that the SFR decreases with increasing galactocentric radius.  Since lower HI column densities are common in the outer discs of nearby galaxies \citep{bigiel10a} and we find lower cloud surface densities in the outer disc of our simulated galaxy, we conclude that one of the most important factors for the SFR is the underlying surface density profile of the disc, consistent with observations and the Kennicutt-Schmidt star formation relation.

To understand the role of external pressure on our clouds, we consider the results of \citet{leroy2015} who used the $v_o$-$\Sigma$ relation to determine whether the clouds in their sample are in virial equilibrium or require pressure confinement to explain their high linewidths.  These authors showed that clouds in the Galactic centre \citep{oka} and the outer Milky Way \citep{hcs01} have high size-linewidth coefficients which can be understood as a reflection of the pressure in the surrounding medium.  While some clouds can be explained with reasonable external pressures, others would require pressures $P/k_{\text{B}}$ $>$ 10$^7$ cm$^{-3}$ K, several decades greater than the typical pressures in the ISM.  Therefore, many of these clouds may be gravitationally unbound.  Although we find in our $v_o$-$\Sigma$ relation shown in Figure~\ref{fg:scaling} that $\sim$ 70 \% of our clouds are gravitationally bound with $\alpha$ $<$ 2 and do not require any pressure confinement, those with $\alpha$ $>$ 2 could be explained with $P/k_{\text{B}}$ $>$ 10$^{4-5}$ cm$^{-3}$ K, consistent with the thermal pressure throughout the disc (Benincasa et al., in preparation).  However, the importance of galactic shear in the disc and the effect of the total angular momentum of the clouds on $\alpha$ and the retrograde population from cloud interactions leads us to conclude that these clouds are in fact unbound.  Also, although we do not find a correlation between the SFR and the virial parameter, we have seen indications that those with the highest SFRs are gravitationally bound, prograde clouds with low axial tilts and 1 $<$ $\alpha$ $<$ 2.  We will explore this further in future work to determine the role of the virial parameter and whether or not the SFR can be strongly linked to the boundedness of molecular clouds.
 
\section*{Acknowledgments}

The authors thank SHARCNET (Shared Hierarchical Academic Research Computing Network), SciNet, and Compute/Calcul Canada, who provided dedicated resources to run these simulations.   
This work was supported by NSERC.  

\bibliographystyle{apj}

\label{lastpage}

\end{document}